\documentclass[twocolumn,showpacs,amsmath,amssymb,prl]{revtex4}

\newcommand{\bsquare}{\hbox{\rule{6pt}{6pt}}}

\begin{document}
\title{Finding a maximally correlated state - Simultaneous Schmidt decomposition of bipartite pure states}
\author{Tohya Hiroshima}
\email{tohya@qci.jst.go.jp}
\author{Masahito Hayashi}
\email{masahito@qci.jst.go.jp}
\affiliation{
Quantum Computation and Information Project, ERATO, Japan Science and Technology Agency,\\
Daini Hongo White Building 201, Hongo 5-28-3, Bunkyo-ku, Tokyo 113-0033, Japan
}
\date{\today}
\begin{abstract}
We consider a bipartite mixed state of the form, 
$\rho =\sum_{\alpha, \beta =1}^{l}a_{\alpha \beta }
\left| \psi _{\alpha }\right\rangle \left\langle \psi_{\beta }\right| $, 
where $\left| \psi _{\alpha}\right\rangle $ are normalized bipartite state vectors, 
and matrix $(a_{\alpha \beta })$ is positive semidefinite.
We provide a necessary and sufficient condition for the state $\rho $ taking the form of maximally correlated states by a local unitary transformation.
More precisely, we give a criterion for simultaneous Schmidt decomposability of 
$\left| \psi _{\alpha }\right\rangle $ for $\alpha =1,2,\cdots ,l$.
Using this criterion, we can judge completely whether or not the state $\rho $ is equivalent to the maximally correlated state, in which the distillable entanglement is given by a simple formula.
For generalized Bell states, this criterion is written as a simple algebraic relation between indices of the states.
We also discuss the local distinguishability of the generalized Bell states that are simultaneously Schmidt decomposable.
\end{abstract}
\pacs{03.67.-a, 03.67.Mn}
\maketitle
The quantum entanglement is well acknowledged to be a physical resource 
in various types of quantum information processing \cite{NC} 
such as quantum cryptography, quantum dense coding, quantum teleportation, and quantum computation.
Therefore, quantifying entanglement is one of the most important issues in quantum information theory.
The entanglement measure for the bipartite pure states is well established.
However, our understanding of entanglement properties of multipartite pure states and general mixed states is still far from satisfactory.
Therefore, exploiting symmetry of entangled states is an effective method to investigate their entanglement properties qualitatively as well as quantitatively \cite{EW,VW01,ADVW}.
Restricted to bipartite mixed states, 
several symmetric states have been proposed and investigated, such as 
Bell diagonal states \cite{BDSW}, Werner states \cite{Werner}, and isotropic states \cite{HH}.
Among them, a maximally correlated state \cite{Rains} on the composite Hilbert space 
${\cal H}_{A}\otimes {\cal H}_{B}$ of the form
\begin{equation} \label{eq:MCS}
\rho _{MC}=\sum_{j,k=1}^{\min \{d_{A},d_{B}\}}\alpha _{jk}
\left|jj\right\rangle \left\langle kk\right|
\end{equation}
has significant entanglement properties.
Here, $d_{A(B)}=\dim {\cal H}_{A(B)}$, and $\left| jj\right\rangle $ denotes 
$\left| j_{A}\right\rangle \otimes \left| j_{B}\right\rangle $ with $\left| j_{A(B)}\right\rangle $ 
being an orthonormal basis in ${\cal H}_{A(B)}$.
A salient feature is that the distillable entanglement $E_{D}$ \cite{BDSW} of the maximally correlated state is given by the following simple formula,
\begin{equation} \label{eq:Distillation}
E_{D}(\rho _{MC})=I_{A}(\rho _{MC})=I_{B}(\rho _{MC}),
\end{equation}
where $I_{A(B)}(\rho )=S(\rho _{A(B)})-S(\rho )$, 
$\rho _{A(B)}={\rm Tr}_{B(A)}\rho $, 
and $S(\rho )=-{\rm Tr}\rho \log _{2}\rho $ 
denotes the von Neumann entropy of $\rho$.
Even in the aforementioned symmetric states other than the maximally correlated state, the formulae of the distillable entanglement are not known.
The formula [Eq.~(\ref{eq:Distillation})] is due to the recently established hashing inequality that gives the lower bound for the distillable entanglement \cite{DW}, 
$E_{D}(\rho )\geq \max \{0,I_{A}(\rho ),I_{B}(\rho )\}$, 
and to the well known fact that the relative entropy of entanglement $E_{R}(\rho ) $ \cite{VPRK,VP} is an upper bound for $E_{D}(\rho ) $ \cite{HHH}.
The relative entropy of entanglement for the maximally correlated state is calculated as 
$E_{R}(\rho _{MC})=I_{A}(\rho _{MC})=I_{B}(\rho _{MC})$ \cite{Rains} so that Eq.~(\ref{eq:Distillation}) follows.

In this paper, we introduce a notion of simultaneous Schmidt decomposition of a set of bipartite pure state vectors.
A necessary and sufficient condition for the simultaneously Schmidt decomposability is given.
We show that a certain class of bipartite mixed states composed of simultaneously Schmidt decomposable vectors can be cast in the maximally correlated states by local unitary transformation.
Because the distillable entanglement of maximally correlated sates is given explicitly, 
the simultaneous Schmidt decomposition is expected to be a useful tool in the distillation problems related with maximally correlated states.
We explore several bipartite mixed states in light of the condition of simultaneous Schmidt decomposability.
Furthermore, for generalized Bell states, 
this condition is shown to be a simple algebraic relation between indices of the states.
Finally we discuss the local distinguishability of the generalized Bell states that are simultaneously Schmidt decomposable.

Let us consider a bipartite mixed state of the form,
\begin{equation} \label{eq:Generalized MCS}
\rho =\sum_{\alpha,\beta =1}^{l}a_{\alpha \beta}
\left| \psi _{\alpha }\right\rangle \left\langle \psi_{\beta }\right|,
\end{equation}
where $\left| \psi _{\alpha }\right\rangle $ are normalized bipartite state vectors in 
${\cal H}_{A}\otimes {\cal H}_{B}$, and matrix $(a_{\alpha \beta})$ is positive semidefinite.
The vectors $\left| \psi _{\alpha }\right\rangle $ do not need to be pairwise orthogonal.
Here, if all $\left| \psi _{\alpha }\right\rangle $ are written as the following form,
\begin{equation} \label{eq:Schmidt}
\left| \psi _{\alpha }\right\rangle =
\sum_{k=1}^{\min \left\{ d_{A},d_{B}\right\} } b_{k}^{(\alpha )}
\left| k_{A}\right\rangle \otimes \left| k_{B}\right\rangle, 
\end{equation}
then the density matrix $\rho $ takes the form of Eq.~(\ref{eq:MCS}) with 
$\alpha _{jk}=\sum_{\alpha ,\beta =1}^{l}a_{\alpha \beta }b_{j}^{(\alpha )}b_{k}^{(\beta )*}$.
Namely, the state $\rho $ is the maximally correlated state.

Even though the coefficients $b_{k}^{(\alpha )} $ in the right hand side of Eq.~(\ref{eq:Schmidt}) are generally complex numbers, we call the decomposition of Eq.~(\ref{eq:Schmidt}) the {\it Schmidt decomposition} of $\left| \psi _{\alpha }\right\rangle $.
In the usual definition \cite{NC}, the expansion coefficients in the Schmidt decomposition must be positive.
However, even when they are complex, their phase factor can be absorbed in the basis vectors making all coefficients real.
Therefore, the definition here is essentially same as the usual one for a single state vector.
However, in this paper, we have occasion to decompose {\it more than one} state vector in the form of Eq.~(\ref{eq:Schmidt}) with common biorthonormal bases so that coefficients $b_{k}^{(\alpha )} $ cannot be real in general even if such a decomposition is possible.
We call the decomposition of this type {\it simultaneous Schmidt decomposition}.

First of all we report the following theorem.

{\em Theorem:}
We associate a $d_{A}\times d_{B}$ matrix,
\begin{equation}
\Psi _{\alpha }=\sum_{j=1}^{d_{A}}\sum_{k=1}^{d_{B}}b_{jk}^{(\alpha )}
\left| j_{A}\right\rangle \left\langle k_{B}\right| ,
\end{equation}
with a bipartite state vector,
\begin{equation}
\left| \psi _{\alpha }\right\rangle
=\sum_{j=1}^{d_{A}}\sum_{k=1}^{d_{B}}b_{jk}^{(\alpha )}
\left| j_{A}\right\rangle \otimes \left| k_{B}\right\rangle .
\end{equation}
If (A) all matrices $\Psi _{\alpha }\Psi _{\beta }^{\dagger }$ 
($\alpha, \beta =1,2,\cdots ,l$) are pairwise commutative, 
then $\Psi _{\alpha}\Psi _{\beta}^{\dagger } $ can be written as
\begin{equation} \label{eq:Expansion1}
\Psi _{\alpha}\Psi _{\beta}^{\dagger }=\sum_{j}\mu _{j}^{(\alpha, \beta )}
\left| v_{j}^{A}\right\rangle \left\langle v_{j}^{A}\right|,
\end{equation}
with $\left| v_{j}^{A}\right\rangle $ being an orthonormal basis in ${\cal H}_{A}$.
Furthermore, if (B) the expansion coefficients in Eq.~(\ref{eq:Expansion1}) satisfy 
$\left| \mu _{j}^{(\alpha,\beta)}\right| ^{2}=\mu _{j}^{(\alpha,\alpha)}\mu _{j}^{(\beta,\beta)}$ 
for $\alpha, \beta =1,2,\cdots ,l$, 
then the vectors $\left| \psi _{\alpha }\right\rangle $ ($\alpha =1,2,\cdots ,l$) are 
simultaneously Schmidt decomposable.
Conversely, if the vectors $\left| \psi _{\alpha}\right\rangle $ ($\alpha =1,2,\cdots ,l$) 
are simultaneously Schmidt decomposable, then conditions (A) and (B) are satisfied.

{\em Proof of Theorem:}
The second part (the converse part) of Theorem is obvious.
The proof of the first part is as follows.
From the first condition (A), 
$\left[ \Psi _{\alpha }\Psi _{\beta }^{\dagger },\Psi _{\beta}\Psi _{\alpha}^{\dagger }\right]=0$, 
we have 
\begin{equation}
\left( \Psi _{\alpha }\Psi _{\beta }^{\dagger }\right)
 \left( \Psi _{\alpha }\Psi _{\beta }^{\dagger }\right) ^{\dagger }=
 \left( \Psi _{\alpha }\Psi _{\beta }^{\dagger }\right) ^{\dagger }
 \left( \Psi _{\alpha }\Psi _{\beta }^{\dagger }\right).
\end{equation}
That is, $\Psi _{\alpha }\Psi _{\beta }^{\dagger }$ 
is a normal matrix and is therefore diagonalizable by an appropriate unitary matrix \cite{HJ}.
Furthermore, all normal matrices $\Psi _{\alpha }\Psi _{\beta }^{\dagger } $ 
are pairwise commutative so that they are diagonalizable by the common unitary matrix \cite{MM}.
Therefore, we have 
\begin{equation} \label{eq:Expansion2}
\Psi _{\alpha }\Psi _{\beta }^{\dagger }=
\sum_{j}\mu _{j}^{(\alpha ,\beta)} \left| v_{j}^{A}\right\rangle \left\langle v_{j}^{A}\right|, 
\end{equation}
for $\alpha, \beta =1,2,\cdots ,l$, where $\left| v_{j}^{A}\right\rangle $ is an orthonormal basis in ${\cal H}_{A}$.
Now, let $P^{(\alpha, \beta)}$ be a projection operator onto 
$W^{(\alpha, \beta)}={\rm supp}\left( \Psi _{\alpha }^{\dagger }\right) \cap 
{\rm supp} \left( \Psi _{\beta }^{\dagger }\right) \neq \emptyset $.
Here, 
${\rm supp}(M) $ denotes $\left\{ \left| v\right\rangle ;M\left| v\right\rangle \neq 0\right\} $.
From Eq.~(\ref{eq:Expansion2}) for $\beta =\alpha $, 
\begin{equation} \label{eq:Psia}
P^{(\alpha, \beta)}\Psi _{\alpha }=
\sum_{j}\sqrt{\mu _{j}^{(\alpha, \alpha)}}P^{(\alpha, \beta)}
\left| v_{j}^{A} \right\rangle \left\langle v_{j}^{(\alpha)}\right|, 
\end{equation}
with $ \left| v_{j}^{(\alpha )}\right\rangle $ 
being an orthonormal basis in ${\cal H}_{B}$.
Similarly,
\begin{equation} \label{eq:Psib}
P^{(\alpha, \beta)}\Psi _{\beta }=
\sum_{j}\sqrt{\mu _{j}^{(\beta, \beta)}}P^{(\alpha, \beta)}
\left| v_{j}^{A}\right\rangle \left\langle v_{j}^{(\beta )}\right|, 
\end{equation}
with $\left| v_{j}^{(\beta)}\right\rangle $ 
being an orthonormal basis in ${\cal H}_{B}$.
From Eqs.~(\ref{eq:Psia}) and (\ref{eq:Psib}), we have
\begin{eqnarray} \label{eq:Psiab}
P^{(\alpha, \beta)}\Psi _{\alpha }\Psi _{\beta }^{\dagger }P^{(\alpha, \beta)} &=&
\sum_{j}\sqrt{\mu_{j}^{(\alpha, \alpha)}\mu _{j}^{(\beta, \beta)}}
\left\langle v_{j}^{(\alpha)}\right| \left.
v_{j}^{(\beta)}\right\rangle  \nonumber \\
&&\times P^{(\alpha, \beta)}\left| v_{j}^{A}\right\rangle 
\left\langle v_{j}^{A}\right| P^{(\alpha, \beta)}.
\end{eqnarray}
From Eqs.~(\ref{eq:Expansion2}) and (\ref{eq:Psiab}), we obtain 
\begin{equation}
\mu _{j}^{(\alpha, \beta)}=
\sqrt{\mu _{j}^{(\alpha, \alpha)}\mu _{j}^{(\beta, \beta)}}
\left\langle v_{j}^{(\alpha)}\right| \left. v_{j}^{(\beta)}\right\rangle. 
\end{equation}
Here, from the second condition (B), 
$\left| \left\langle v_{j}^{(\alpha)}\right| \left. v_{j}^{(\beta)}\right\rangle \right| =1$, i.e., $\left| v_{j}^{(\beta)} \right\rangle $ 
is the same as 
$\left| v_{j}^{(\alpha)} \right\rangle $ 
apart from the phase factor $e^{\sqrt{-1}\phi _{\beta}}$.
Therefore, vectors 
$P^{(\alpha ,\beta )}\left| \psi _{\alpha }\right\rangle $ and 
$P^{(\alpha ,\beta )}\left| \psi _{\beta }\right\rangle $ 
are expanded with the common biorthogonal basis, 
$\left| v_{j}^{A}\right\rangle \otimes \left| v_{j}^{(\alpha )}\right\rangle $.
If ${\rm supp}\left( \Psi _{\alpha}^{\dagger }\right) \cap 
{\rm supp}\left(\Psi _{\beta}^{\dagger }\right) =\emptyset $, 
then $\left| v_{j}^{(\alpha)}\right\rangle $ 
with $\left| v_{j}^{A} \right\rangle \in {\rm supp}\left( \Psi _{\alpha }^{\dagger }\right) $ 
and $\left| v_{j^{\prime }}^{(\beta)} \right\rangle $ 
with $\left| v_{j^{\prime }}^{A} \right\rangle \in {\rm supp}\left( \Psi _{\beta }^{\dagger }\right) $ 
must be orthogonal to each other; otherwise 
$\Psi _{\alpha }\Psi _{\beta }^{\dagger }$ 
would not be the form of Eq.~(\ref{eq:Expansion2}).
Thus, all $\left| \psi _{\alpha}\right\rangle $ 
are written as the simultaneously Schmidt decomposed form.
This completes the proof of Theorem. \hspace*{\fill} \bsquare

In the following, we examine the above described conditions with some examples.
The first example is a set of two state vectors in a $4\otimes 4$ system: 
$\left| \psi _{1}\right\rangle =
\left( \left| 00\right\rangle +\left| 12\right\rangle +\left| 21\right\rangle \right) /\sqrt{3}$ 
and 
$\left| \psi _{2}\right\rangle =
\left( \left| 13\right\rangle +\left| 21\right\rangle +\left| 33\right\rangle \right) /\sqrt{3}$.
We can readily see that these two vectors cannot be written as the simultaneously Schmidt decomposed form because 
$\left| 12\right\rangle $ in $\left| \psi _{1}\right\rangle $ and $\left| 13\right\rangle $ 
in $\left| \psi _{2}\right\rangle $ cannot be cast in the same form by a unitary transformation.
A direct computation yields
$\Psi _{1}\Psi _{1}^{\dagger }=
\left( \left| 0_{A}\right\rangle \left\langle 0_{A}\right| +
\left| 1_{A}\right\rangle \left\langle 1_{A}\right| +
\left| 2_{A}\right\rangle \left\langle 2_{A}\right| \right)/3$, 
$\Psi _{2}\Psi _{2}^{\dagger }=
\left( \left| 1_{A}\right\rangle \left\langle 1_{A}\right| +
\left| 2_{A}\right\rangle \left\langle 2_{A}\right| +
\left| 3_{A}\right\rangle \left\langle 3_{A}\right| \right)/3$, and
$\Psi _{1}\Psi _{2}^{\dagger }=\Psi _{2}\Psi _{1}^{\dagger }=
\left( \left| 2_{A}\right\rangle \left\langle 2_{A}\right| \right)/3$.
Therefore, condition (B) is violated 
(
$\left| \mu _{j}^{(1,2)}\right| ^{2}=
\left| \mu _{j}^{(2,1)}\right| ^{2}=0\neq \mu _{j}^{(1,1)}\mu _{j}^{(2,2)}=1/9$ 
for 
$\left| j_{A}\right\rangle =\left| 1_{A}\right\rangle $
) even though condition (A) is still satisfied.

The second example is again a set of two state vectors in a $4\otimes 4$ system \cite{EFPPW}: 
$\left| \psi _{1}\right\rangle =\left( \left| 11\right\rangle -\left|
12\right\rangle -\left| 21\right\rangle +\left| 22\right\rangle \right) /2$ 
and 
$\left| \psi _{2}\right\rangle =\left( 2\left| 00\right\rangle +\left|
11\right\rangle +\left| 12\right\rangle +\left| 21\right\rangle +\left|
22\right\rangle +2\left| 33\right\rangle \right) /\sqrt{12}$.
Eisert {\it et al.} showed by explicit calculations that 
the distillable entanglement of the statistical mixture of these two pure states, 
$\sigma =(1/4)\left| \psi _{1}\right\rangle \left\langle \psi _{1}\right| 
+(3/4)\left| \psi _{2}\right\rangle \left\langle \psi _{2}\right|$ 
is exactly the relative entropy of entanglement of $\sigma $ \cite{EFPPW}.
The state $\sigma $ is indeed a maximally correlated state.
It is easy to show 
$ \Psi _{1}\Psi _{1}^{\dagger }=\left| v_{4}^{A}\right\rangle \left\langle v_{4}^{A}\right| $, 
$ \Psi _{2}\Psi _{2}^{\dagger }=\left( \left|
v_{1}^{A}\right\rangle \left\langle v_{1}^{A}\right| +\left|
v_{2}^{A}\right\rangle \left\langle v_{2}^{A}\right| +\left|
v_{3}^{A}\right\rangle \left\langle v_{3}^{A}\right| \right)/3 $, 
and 
$ \Psi _{1}\Psi _{2}^{\dagger }=\Psi _{2}\Psi _{1}^{\dagger }=0 $ 
with 
$\left| v_{1}^{A}\right\rangle =\left| 0_{A}\right\rangle $, 
$\left| v_{2}^{A}\right\rangle =\left| 3_{A}\right\rangle $, 
$\left| v_{3}^{A}\right\rangle =
\left( \left| 1_{A}\right\rangle +\left| 2_{A}\right\rangle \right) /\sqrt{2}$, 
and 
$\left| v_{4}^{A}\right\rangle =
\left( \left| 1_{A}\right\rangle -\left| 2_{A}\right\rangle \right) /\sqrt{2}$.
Therefore, conditions (A) and (B) are satisfied, and the vectors 
$\left| \psi _{1}\right\rangle $ 
and 
$\left| \psi _{2}\right\rangle $ 
can be written as the simultaneously Schmidt decomposed form: 
$\left| \psi _{1}\right\rangle =
\left| v_{4}^{A}\right\rangle \otimes \left| v_{4}^{B}\right\rangle $
and 
$\left| \psi _{2}\right\rangle =\left( 1/\sqrt{3}\right)
\sum_{j=1}^{3}\left| v_{j}^{A}\right\rangle \otimes \left|v_{j}^{B}\right\rangle $, 
where 
$\left| v_{1}^{B}\right\rangle =\left| 0_{B}\right\rangle $, 
$\left| v_{2}^{B}\right\rangle =\left| 3_{B}\right\rangle $, 
$\left| v_{3}^{B}\right\rangle =
\left( \left| 1_{B}\right\rangle +\left| 2_{B}\right\rangle \right) /\sqrt{2}$, 
and 
$\left| v_{4}^{B}\right\rangle =
\left( \left| 1_{B}\right\rangle -\left| 2_{B}\right\rangle \right) /\sqrt{2}$.

A generalized Bell diagonal state or, more generally, 
a mixed state of the form of Eq.~(\ref{eq:Generalized MCS}) with 
$\left| \psi _{\alpha }\right\rangle $ ($\alpha =1,2,\cdots ,l$) 
being generalized Bell states is of particular interest.
The generalized Bell states in a $d\otimes d$ system are defined as
\begin{equation} \label{eq:Bell}
\left| \psi _{nm}^{(d)}\right\rangle =\left( Z^{n} \otimes
X^{-m}\right) \left| \psi _{+}^{(d)}\right\rangle ,
\end{equation}
for $n,m=0,1,\cdots ,d-1$, where 
$\left| \psi _{+}^{(d)}\right\rangle =d^{-1/2}\sum_{k=0}^{d-1}\left| k\right\rangle \otimes \left| k\right\rangle 
$, and unitary matrices $X$ and $Z$ are defined by 
$X\left| k\right\rangle=\left| k-1\,(\mathrm{mod}\,d)\right\rangle $ and 
$Z\left| k\right\rangle=\omega _{d}^{k}\left| k\right\rangle $ for $k=0,1,\cdots ,d-1$ with 
$\omega_{d}=\exp \left( 2\pi \sqrt{-1}/d\right) $ \cite{Hamada}.
These vectors are pairwise orthogonal maximally entangled states.
It is easy to show the following relation, 
\begin{equation} \label{eq:Weyl}
XZ=\omega _{d}ZX.
\end{equation}
The associated matrices are calculated as 
\begin{equation} \label{eq:BellMatrix}
\Psi _{nm}^{(d)}=\frac{1}{\sqrt{d}}Z^{n}\sum_{k=0}^{d-1}\left|
k\right\rangle \left\langle k\right| X^{m}=\frac{1}{\sqrt{d}}Z^{n}X^{m}.
\end{equation}
From Eq.~(\ref{eq:BellMatrix}), we have
\begin{equation} \label{eq:Identity}
\Psi _{nm}^{(d)}\Psi _{nm}^{(d)\dagger }=
\Psi _{nm}^{(d)\dagger }\Psi _{nm}^{(d)}=\frac{1}{d}{\bf I},
\end{equation}
where ${\bf I}$ is the $d\times d$ identity matrix.

Condition (B) is always satisfied for a given set of the generalized Bell states 
$\left| \psi _{n_{\alpha }m_{\alpha }}^{(d)}\right\rangle $ ($\alpha =1,2,\cdots ,l$) 
provided that condition (A) is satisfied for these vectors.
This can be seen as follows.
Suppose that condition (A) is satisfied.
Then, 
$ \Psi _{n_{\alpha }m_{\alpha }}^{(d)}\Psi _{n_{\beta }m_{\beta }}^{(d)\dagger}=
\sum_{j=1}^{d}\lambda _{j} \left| v_{j}^{A} \right\rangle \left\langle v_{j}^{A}\right| $, 
where $ \left| v_{j}^{A} \right\rangle $ is an orthonornal basis in ${\cal H}_{A} $.
Therefore,
\begin{equation} \label{eq:Squared}
\Psi _{n_{\alpha }m_{\alpha }}^{(d)}\Psi _{n_{\beta }m_{\beta }}^{(d)\dagger }
\left( \Psi _{n_{\alpha }m_{\alpha }}^{(d)}
\Psi _{n_{\beta }m_{\beta }}^{(d)\dagger }\right) ^{\dagger }=
\sum_{j=1}^{d} \left| \lambda _{j} \right| ^{2} 
\left| v_{j}^{A} \right\rangle \left\langle v_{j}^{A}\right|. 
\end{equation}
Because of Eq.~(\ref{eq:Identity}), the left-hand side of Eq.~(\ref{eq:Squared}) is 
${\bf I}/d^{2}$.
Thus $\left| \lambda _{j} \right| =1/d$.
Furthermore, due to Eq.~(\ref{eq:Identity}), 
\begin{equation}
\Psi _{n_{\alpha }m_{\alpha }}^{(d)}\Psi _{n_{\alpha }m_{\alpha
}}^{(d)\dagger }=\Psi _{n_{\beta }m_{\beta }}^{(d)}\Psi _{n_{\beta }m_{\beta
}}^{(d)\dagger }=\sum_{j=1}^{d} \frac{1}{d} \left| v_{j}^{A}\right\rangle
\left\langle v_{j}^{A}\right|.
\end{equation}
Consequently, condition (B) is satisfied.
We thus have to check only condition (A) to find a set of vectors 
$\left| \psi _{n_{\alpha }m_{\alpha }}^{(d)}\right\rangle $ ($\alpha =1,2,\cdots ,l$) 
that can be written as a simultaneously Schmidt decomposed form.

Condition (A) is rewritten as an algebraic relation between pairs of indices $(n,m)$ of 
$\left| \psi _{nm}^{(d)}\right\rangle $.
Using Eqs.~(\ref{eq:Weyl}) and (\ref{eq:BellMatrix}), we obtain
\begin{eqnarray}
&&\left[ \Psi _{n_{\alpha }m_{\alpha }}^{(d)}\Psi _{n_{\beta }m_{\beta
}}^{(d)\dagger },\Psi _{n_{\gamma }m_{\gamma }}^{(d)}\Psi _{n_{\delta
}m_{\delta }}^{(d)\dagger }\right]   \nonumber \\
&=&d^{-2}\omega _{d}^{-n_{\alpha }(m_{\alpha }-m_{\beta })-n_{\gamma
}(m_{\gamma }-m_{\delta })}  \nonumber \\
&&\times \left[ \omega _{d}^{-(n_{\alpha }-n_{\beta })(m_{\gamma }-m_{\delta
})}-\omega _{d}^{-(n_{\gamma }-n_{\delta })(m_{\alpha }-m_{\beta })}\right] 
\nonumber \\
&&\times X^{m_{\alpha }-m_{\beta }+m_{\gamma }-m_{\delta }}Z^{n_{\alpha
}-n_{\beta }+n_{\gamma }-n_{\delta }}.
\end{eqnarray}
Therefore, 
$\Psi _{n_{\alpha }m_{\alpha }}^{(d)}\Psi _{n_{\beta }m_{\beta }}^{(d)\dagger }$ 
and 
$\Psi _{n_{\gamma }m_{\gamma }}^{(d)}\Psi _{n_{\delta }m_{\delta }}^{(d)\dagger }$ 
commute each other if and only if 
$(n_{\alpha }-n_{\beta })(m_{\gamma }-m_{\delta })=
(n_{\gamma }-n_{\delta })(m_{\alpha }-m_{\beta })$ 
(${\rm mod}\,d$) holds or, equivalently, 
there exist integers $p$ and $q$ ($p\neq 0$ or $q\neq 0$) satisfying 
$ p(n_{\alpha }-n_{\beta })+q(m_{\alpha }-m_{\beta })=0 $ (${\rm mod}\,d$) 
and 
$ p(n_{\gamma }-n_{\delta })+q(m_{\gamma }-m_{\delta })=0 $ (${\rm mod}\,d$).
Therefore, vectors 
$\left| \psi _{n_{\alpha }m_{\alpha }}^{(d)}\right\rangle $ ($\alpha =1,2,\cdots ,l$) 
are simultaneously Schmidt decomposable if and only if 
for any two elements $(n,m)$ and $(n^{\prime },m^{\prime })$ in 
$\left\{ (n_{\alpha },m_{\alpha })\right\} _{\alpha =1}^{l}$, 
there exist integers $p$ and $q$ ($p\neq 0$ or $q\neq 0$) 
satisfying
\begin{equation} \label{eq:algebraic}
p(n-n^{\prime })+q(m-m^{\prime })=0 \quad ({\rm mod}\,d).
\end{equation}
This is equivalent to the following condition (${\rm A}^{\prime }$): 
There exist integers $p$, $q$, and $r$ ($p\neq 0$ or $q\neq 0$) satisfying
\begin{equation}
pn_{\alpha }+qm_{\alpha }=r \quad ({\rm mod}\,d)
\end{equation}
for all $\alpha =1,2,\cdots ,l$.

From the above condition, we can find a set of generalized Bell states 
that are simultaneously Schmidt decomposable.
For example, any two independent generalized Bell states are simultaneously Schmidt decomposable.
This is because it is always possible to find integers $p$ and $q$ ($p\neq 0$ or $q\neq 0$) 
satisfying Eq.~(\ref{eq:algebraic}) for any two index pairs 
$(n,m)$ and $(n^{\prime },m^{\prime })$.
Therefore, the generalized Bell diagonal states of rank 2, 
$\rho =\lambda \left| \psi _{nm}^{(d)}\right\rangle \left\langle \psi _{nm}^{(d)}\right| +
(1-\lambda )\left| \psi _{n^{^{\prime }}m^{^{\prime }}}^{(d)}\right\rangle 
\left\langle \psi _{n^{^{\prime }}m^{^{\prime }}}^{(d)}\right| $ 
with $0<\lambda <1$ take the form of maximally correlated states by local unitary transformation.
The distillable entanglement of this state is therefore given by 
$E_{D}(\rho )=\log _{2}d-S(\rho ) $.
This is the generalization of the known result that the distillable entanglement of 
Bell diagonal states of rank 2 is given by $1-S(\rho )$ \cite{BDSW}.
More generally, the mixed state 
$ \rho =\sum_{\alpha, \beta =1}^{2}a_{\alpha \beta }
\left| \psi _{n_{\alpha }m_{\alpha }}^{(d)}\right\rangle
\left\langle \psi _{n_{\beta }m_{\beta }}^{(d)}\right| $ 
also takes the form of maximally correlated states by local unitary transformation.

More than two vectors taken from the set of generalized Bell states of Eq.~(\ref{eq:Bell}) 
do not always satisfy condition (A) or (${\rm A}^{\prime }$).
For example, in a $3\otimes 3$ system, 
we have only 12 sets of three independent vectors that satisfy condition (${\rm A}^{\prime }$).
In a $4\otimes 4$ system, we have 112 sets of three independent vectors 
and 28 sets of four independent vectors, both of which satisfy condition (${\rm A}^{\prime }$).
In a $5\otimes 5$ system, we have 300 sets of three independent vectors, 
150 sets of four independent vectors, 
and 30 sets of five independent vectors, all of which satisfy condition (${\rm A}^{\prime }$).
These numbers are easily computed by utilizing commercially available mathematics software.
In particular, the following special sets of index pairs, 
$\left\{ (n,fn+g \,({\rm mod}\,d))\right\} _{n=0}^{d-1}$ 
and 
$\left\{ (fm+g \,({\rm mod}\,d),m)\right\} _{m=0}^{d-1}$ 
satisfy condition (${\rm A}^{\prime }$), where $f$ and $g$ are arbitrary integers.
Therefore, for example, the statistical mixture of 
$\left| \psi _{nm}^{(d)}\right\rangle \left\langle \psi_{nm}^{(d)}\right| $ 
($m=0,1,\cdots ,d-1$) 
or that of 
$\left| \psi _{nm}^{(d)}\right\rangle \left\langle \psi_{nm}^{(d)}\right| $ 
($n=0,1,\cdots ,d-1$) 
takes the form of maximally correlated states.
This has been previously pointed out in \cite{VW03,VWW}.
More than $d$ generalized Bell states cannot be simultaneously Schmidt decomposable.
If this is the case, we could construct a maximally correlated state of rank $d^{\prime }(>d)$, which contradicts the obvious fact that the rank of the maximally correlated states cannot exceed $d$.

It should be noted that the generalized Bell states which are simultaneously 
Schmidt decomposable can be distinguished deterministically 
by local operations and classical communication (LOCC) \cite{VSPM}.
This can be seen as follows \cite{Fan}.
Suppose that Alice and Bob share a set of simultaneously Schmidt decomposable vectors 
$\left| \psi _{n_{\alpha }m_{\alpha }}^{(d)}\right\rangle $ ($\alpha =1,2,\cdots ,l$).
They can always find a local unitary transformation $U_{A}\otimes U_{B}$ such that
\begin{eqnarray} \label{eq:distinguish}
\left( U_{A}\otimes U_{B}\right) \left| \psi _{n_{\alpha }m_{\alpha
}}^{(d)}\right\rangle  &=&\frac{1}{\sqrt{d}}\sum_{k=0}^{d-1}\omega
_{d}^{kr_{\alpha }}\left| k\right\rangle \otimes \left| k\right\rangle  \nonumber \\
&=&\left( Z^{r_{\alpha }}\otimes \mathbf{I}\right) \left| \psi
_{+}^{(d)}\right\rangle. 
\end{eqnarray}
In Eq.~(\ref{eq:distinguish}), all $r_{\alpha }$ are different; 
otherwise all of the vectors given by Eq.~(\ref{eq:distinguish}) would not be pairwise orthogonal.
Next, Alice and Bob perform unitary transformations $H_{0}$ and $H_{0}^{*}$, respectively, 
where $H_{0}$ is defined by $\left( H_{0}\right) _{jk}=d^{-1/2}\omega _{d}^{-jk}$.
Therefore, vectors $\left| \psi _{n_{\alpha }m_{\alpha }}^{(d)}\right\rangle $ 
are transformed into 
$\left( H_{0}Z^{r_{\alpha }}\otimes H_{0}^{*}\right) \left| \psi _{+}^{(d)}\right\rangle $, 
which can be written as 
$\left( H_{0}Z^{r_{\alpha }}H_{0}^{\dagger }\otimes \mathbf{I}\right) \left|
\psi _{+}^{(d)}\right\rangle =\left( X^{-r_{\alpha }}\otimes \mathbf{I}%
\right) \left| \psi _{+}^{(d)}\right\rangle $.
All of these vectors are distinguishable by projecting measurements in the basis 
$\left| k \right\rangle $ ($k=0,1,\cdots ,d-1$) on both sides 
followed by classical communication.
Therefore, condition (${\rm A}^{\prime }$) provides a sufficient condition for generalized Bell states being deterministically distinguished by LOCC.
Note that this condition is generally weaker than that given in \cite{GKRS}.

In conclusion, we introduced the notion of simultaneous Schmidt decomposition and gave the necessary and sufficient condition or criterion for simultaneous Schmidt decomposability.
Using this criterion, we can judge completely whether or not a bipartite mixed state [Eq.~(\ref{eq:Generalized MCS})] is equivalent to the maximally correlated state.
For generalized Bell states, a simple algebraic criterion was found for simultaneous Schmidt decomposability.
It is also a sufficient condition for deterministic LOCC distinguishability of the generalized Bells states.

{\it Acknowledgements}: The authors are grateful to Dr. Mitsuru Hamada for his useful discussions and to Professor Hiroshi Imai for his support and encouragement.


\end{document}